\documentclass[letter]{aa}  

\usepackage{graphicx}
\usepackage{lscape}
\usepackage{hyperref}
\usepackage{txfonts}
\usepackage{natbib}
\bibpunct{(}{)}{;}{a}{}{,} 

\begin{document}

   \title{OGLE\,GD-CEP-0516: the most metal-poor lithium-rich Galactic Cepheid.\thanks{Based on observations European Southern Observatory programs P105.20MX.001}}


\author{G. Catanzaro \inst{1}
          \and
          V. Ripepi \inst{2}
          \and
          M. Salaris \inst{3,4}
          \and
          E. Trentin \inst{5,6,2}
 }

\institute{INAF-Osservatorio Astrofisico di Catania, Via S.Sofia 78, 95123, Catania, Italy \\ \email{giovanni.catanzaro@inaf.it}
             \and
             INAF-Osservatorio Astronomico di Capodimonte, Salita Moiariello 16, 80131, Naples, Italy\\  \email{vincenzo.ripepi@inaf.it}
             \and
             Astrophysics Research Institute, Liverpool John Moores University, 146 Brownlow Hill, Liverpool L3 5RF, UK\\ \email{m.salaris@ljmu.ac.uk}
             \and
             INAF - Osservatorio Astronomico di Abruzzo, Via M. Maggini, s/n, I-64100 Teramo, Italy
             \and
             Leibniz-Institut f\"ur Astrophysik Potsdam (AIP), An der Sternwarte 16, D-14482 Potsdam, Germany \\ \email{etrentin@aip.de}
             \and
             Institut für Physik und Astronomie, Universität Potsdam, Haus 28, Karl-Liebknecht-Str. 24/25, D-14476 Golm (Potsdam), Germany\\
             }

   \date{}

 
  \abstract
   {Classical Cepheids (DCEPs) are important astrophysical objects not only as standard candles for the determination of the cosmic distance ladder but also as a test-bed for the stellar evolution theory, thanks to the connection between their pulsation (periods, amplitudes)  and stellar (luminosity, mass, effective temperature, metallicity) parameters.}   
   {We aim at unveiling the nature of the Galactic DCEP OGLE-GD-CEP-0516 and other DCEPs showing enhanced abundance of lithium in their atmospheres.}
   {We have collected high-resolution spectrum for OGLE-GD-CEP-0516 with UVES@VLT. Accurate stellar parameters: effective temperature, gravity, micro- and macro-turbulence, radial velocity, and metal abundances were measured for this star, by using spectral synthesis techniques based on LTE plane-parallel atmospheric model.}
   {We measured a chemical pattern with most elements under-abundant compared with the Sun, i.e. [Fe/H]\,=\,$-$0.54\,$\pm$\,0.16~dex, [C/H]\,=\,$-$0.45\,$\pm$\,0.05~dex, or [Mg/H]\,=\,$-$0.40\,$\pm$\,0.16~dex among others. In particular, we measured a lithium abundance A(Li)\,=\,3.06\,$\pm$\,0.10~dex for OGLE-GD-CEP-0516, which makes this object the sixth Li-rich object among the Milky Way DCEPs.}
   {Our results favour the scenario in which the six Galactic Li-rich DCEPs are first-crossing the instability strip having had slowly-rotating progenitors during their main sequence phase. This study explored the link between lithium abundance and the pulsation period in classical Cepheids. It found that brighter Cepheids, indicative of higher mass, show enhanced lithium abundance, contrary to predictions from evolutionary models considering rotation. Additionally, an analysis of lithium abundance versus [Fe/H] revealed a lack of significant correlation, contradicting expectations from galactic chemical evolution (GCE) models.}


   \maketitle
%
\section{Introduction}


Classical Cepheids (DCEPs) are the most important standard candle for the extragalactic distance scale and are powerful tracers of the young ($\sim$ 50--300 Myr) populations inside a galaxy, including the Milky Way. Moreover, observations of DCEPs are also crucial to further our understanding of the physical mechanisms governing their evolution and pulsation. In this context, the rare lithium-rich DCEPs subclass is of particular interest.    
Indeed, only five DCEPs showing enhanced lithium abundance (through the detection of the \ion{Li}{I} 6707.766 \AA) have been discovered in the Galaxy so far \citep{2011AJ....142..136L,2016MNRAS.460.2077K,2019MNRAS.488.3211K,catanzaro2020} and an additional one was detected in the Large Magellanic Cloud \citep[LMC,][]{1992ApJS...79..303L}. All these objects show a surface lithium abundance  A(Li)$\sim$3.0 dex, in contrast to the majority of the Galactic DCEPs,  which show A(Li)$<$1.2 dex \citep{2011AJ....142..136L}. This discovery was surprising, as Li is expected to be depleted by proton-capture after the first dredge-up (1DU) occurring at the beginning of the Red Giant Branch (RGB) phase \citep{1967ARA&A...5..571I}. A natural explanation is that these  DCEPs are at their first crossing of the IS and their envelopes do not show the signature of nuclear processes that occurred during the Main Sequence (MS) phase. Indeed, according to \citet{2019MNRAS.488.3211K}, at least three of the four Milky Way (MW) Li-rich DCEPs also show abundances of the CNO species which are consistent with the solar values, i.e. not processed by the CN-cycle. 
However,  the 1DU is not the only phenomenon capable of depleting lithium. Rotational mixing can reduce the lithium abundance by a factor of a hundred in a fraction of the MS lifetime, for sufficiently fast rotating MS stars \citep[e.g.][]{2011A&A...530A.115B}. This would then explain the scant number of Li-rich DCEPs. Indeed, as noted by \citet{2019MNRAS.488.3211K}, about 80\% of the DCEPs expected to be at their first crossing (about 5\% of the total) are Li-depleted. Therefore it can be hypothesized that the progenitors of the Li-rich and Li-depleted DCEPs (B stars) were, respectively, slow and fast rotators during their MS evolution. It is known that a fraction ($\sim$15\%) of the B-stars show v$\sin i <$\, 20 km s$^{-1}$ \citep{2010ApJ...722..605H}, while the  large majority rotates much faster. It is thought that the slow rotators lose most of their angular momentum on the MS due to stellar winds enhanced by the rotation itself \citep{2000ARA&A..38..143M}. Therefore, upon becoming DCEPs they show the moderate rotational velocities typical of these stars. \par
An additional feature of the Li-rich DCEPs is that they most frequently are multi-mode pulsators. Among the five MW Li-rich DCEPs, ASAS J075842-2536.1, ASAS J131714-6605.0 and V363\,Cas pulsate in the first and second overtone (DCEP\_1O2O), V371\,Per pulsates in the fundamental and first overtone (DCEP\_F1O), whereas V1033\,Cyg is only a fundamental mode (DCEP\_F) pulsator. 
According to  \citet{2019MNRAS.488.3211K}, multi-mode DCEPs have a less efficient mixing in their envelope than DCEP\_F, hence would preferentially tend to retain their Li. 

Even if other more complex processes can address the presence of lithium in DCEPs \citep[see][for a detailed discussion]{2019MNRAS.488.3211K}, the basic mechanism to explain Li-rich DCEPs is their transit through the IS at the first crossing. This occurrence can be verified by measuring the rate of period change due to their evolution along the Hertzsprung-Russell Diagram (HRD), as the period is expected to increase at the first and third crossing, and decrease at the second one \citep[see e.g.][]{2006PASP..118..410T}. The data available allowed \cite{2019MNRAS.488.3211K} to detect a quick period change in V1033\,Cyg and \citet{catanzaro2020} in V363\,Cas whereas they were insufficient to detect period changes in the other three MW DCEPs.  


In the course of a large project devoted to measuring the metallicity dependence of DCEPs period-luminosity relations dubbed C-MetaLL\footnote{https://sites.google.com/inaf.it/c-metall/home} \citep[Cepheid - Metallicity in the Leavitt Law, see ][]{ripepi2021,trentin2022}, we obtained high-resolution spectroscopy for hundreds of Galactic DCEPs. As a by-product of the C-MetaLL projects, we searched systematically all the targets for the presence of a deep \ion{Li}{I} 6707.766 {\AA} line in their spectra. In this letter, we report the successful detection of a such lithium line in the  Galactic DCEP OGLE-GD-CEP-0516. 

\begin{table}
\centering
\caption{Main characteristics of OGLE-GD-CEP-0516.}
\label{tab:star}
\setlength{\tabcolsep}{3.5pt}
\begin{tabular}{ccccc}
\hline
\hline            
\noalign{\smallskip}
{\it Gaia DR3} ID & P & Mode  & G & A$_V$ \\
\noalign{\smallskip}
       & days &  & mag & mag \\
\noalign{\smallskip}
\hline            
\noalign{\smallskip}       
5255256669866274816 & 0.394959 & 1O/2O & 12.462 & 1.446$^{a}$ \\

\hline                                      \end{tabular}
\tablefoot{~a =  \citet{2019A&A...628A..94A}}
\end{table}

\section{Spectroscopic observations and data analysis}
\label{sect:spect}

We collected high-resolution spectroscopy for OGLE-GD-CEP-0516 using the Ultraviolet and Visual Echelle Spectrograph (UVES)\footnote{https://www.eso.org/sci/facilities/paranal/instruments/uves.html}) instrument, operated by the European Southern Observatory (ESO) and attached to the Unit Telescope 2 of the Very Large Telescope (VLT), placed at Paranal (Chile). The spectrum covers the wavelength range between $\lambda \lambda$ 4790 to 6800 {\AA}, with a spectral resolution R=47\,000. The signal-to-noise ratio (SNR) varies from 80 to 100. The main characteristics of OGLE-GD-CEP-0516 are summarised in Table~\ref{tab:star}.

The spectrum reduction, which included bias subtraction, spectrum extraction, flat fielding, and wavelength calibration, was performed using the ESO reduction pipeline. 
The radial velocity was measured by cross-correlating the observed spectrum with a synthetic template, using the Image Reduction and Analysis Facility (IRAF) task {\tt FXCOR} and excluding Balmer lines as well as wavelength ranges containing telluric lines. The IRAF package {\tt RVCORRECT} was adopted to determine the heliocentric velocity by correcting the spectrum for the Earth’s motion.

\begin{figure*}
\centering
\includegraphics[width=13cm]{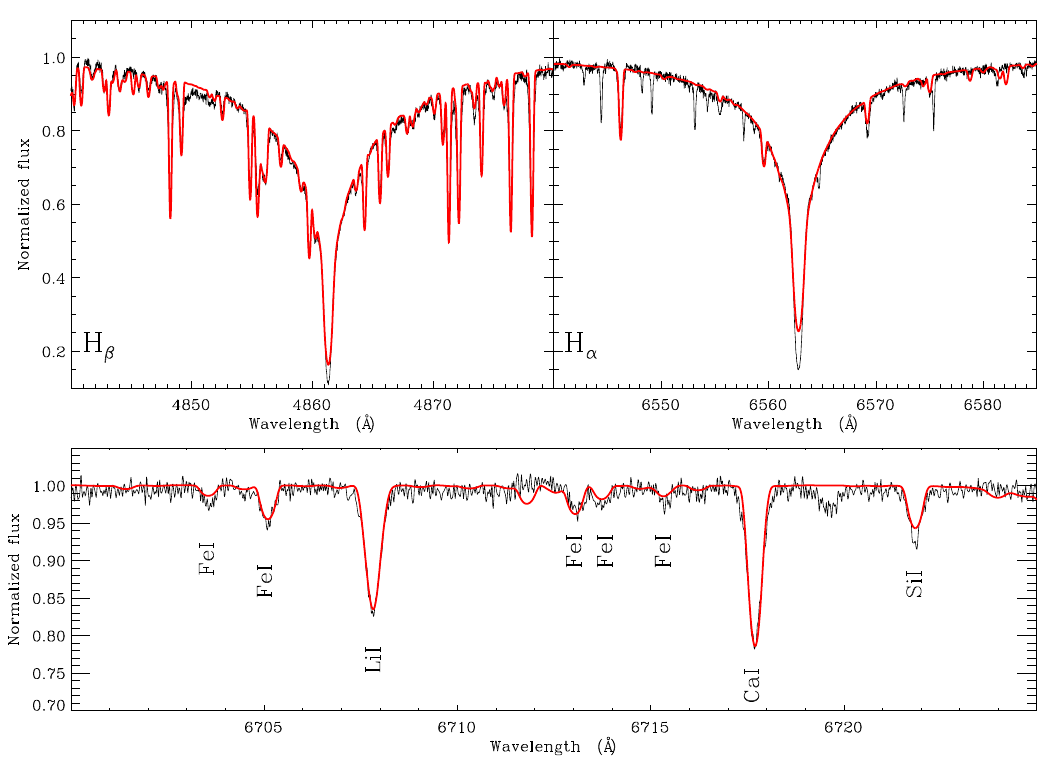}
\caption{Comparison between the spectrum of  OGLE-GD-CEP-0516 obtained (black line) and the synthetic spectrum (red line) in three main spectral intervals centred, respectively, on H$\beta$ (top-left panel), H$\alpha$ (top-right panel) and the \ion{Li}{I} line at 6707.766 {\AA} (bottom panel). The spectral features not reproduced with synthetic spectrum in the wing of H$\alpha$ are telluric lines. The lithium line has been reproduced with A(Li)\,=\,3.06\,$\pm$\,0.10~dex.}
\label{spectra}
\end{figure*}

All the important quantities characterizing the stellar atmosphere, such as effective temperature (T$_{\rm eff}$), surface gravity ($\log g$), microturbulent ($\xi$), and lines broadening contributions (v$_{br}$) due to the combined effects of v $\sin i$ and macroturbulent velocity (that is a dominant contribution in DCEP), have been already derived in \citet{trentin2022}. These quantities are listed in Table~\ref{table_sum_spectro}.

\begin{table*}
\centering
\caption{Atmospheric parameters adopted for OGLE-GD-CEP-0516. We list adopted effective temperature, gravity, microturbulence, line broadening, projected rotational velocity, radial-tangential macroturbulence, heliocentric Julian day of the middle exposure, and radial velocities.}
\label{table_sum_spectro}
\begin{tabular}{ccccccccc}
\hline \hline
 T$_{\rm eff}$   &     $\log g$    &  $\xi$          &   v$_{br}$       & v$\sin i$      & $\Theta_{RT}$  &  HJD       &   v$_{\rm rad}$        \\
      (K)        &                 &  (km s$^{-1}$)  & (km s$^{-1}$) & (km s$^{-1}$)  & (km s$^{-1}$)  & (2450000.0+)     &   (km s$^{-1}$)        \\
\hline                                      
6400 $\pm$ 150   & 1.5 $\pm$ 0.1   &  2.4 $\pm$ 0.2  & 13 $\pm$ 1    & 9.2 $\pm$ 0.6  & 12.3 $\pm$ 2.4 &  9217.6507 &  $-$9.0 $\pm$ 0.2    \\
\hline                     
\end{tabular}
\end{table*}

As a check of these parameters (principally T$_{eff}$ and $\log g$), we reproduced the observed spectral energy distribution (SED) with the synthetic flux computed using the plane parallel local thermodynamic equilibrium (LTE) atmosphere models computed using the ATLAS9 code \citep{1993ASPC...44...87K}. The observed flux was retrieved from the Virtual Observatory SED Analyzer \citep[VOSA][]{2008A&A...492..277B} and corrected for reddening adopting A$_V$\,=\,1.446 mag \citep{2019A&A...628A..94A} and the \citet{1999PASP..111...63F} extinction law. 
The comparison between the observed and the theoretical SEDs is satisfactory, as shown in Fig.~\ref{sed}. The SED can be used to estimate the bolometric luminosity $L_{bol}$ of OGLE\,GD-CEP-0516, provided we know its distance. To this aim we adopted the distance obtained by \citet{Bailer2021} based on $Gaia$ Early Data Release 3 \citep[EDR3][]{Gaia2016,GaiaBrown2021}, obtaining a final value of $L_{bol}$\,=\,140\,$\pm$\,5 $L_\odot$.

As an additional check regarding the line broadening value reported in \citet{trentin2022}, we adopted the code {\tt iacob-broad} described in \citet{2014A&A...562A.135S} to disentangle the effects of the rotational velocity to that of macroturbulence. In this code, the authors assumed a radial-tangential definition of the macroturbulence profile \citep[see][for a detailed description]{2008oasp.book.....G}. In a few words, we derived the macroturbulent velocity from the goodness-of-fit method (using a $\chi^2$ formalism) when the v\,$\sin i$ is fixed to the value corresponding to the first zero of the Fast Fourier Transform (FFT) of a chosen line profile. For this calculation, we have chosen five spectral lines well isolated in the spectrum and for which signal-to-noise on both sides of the line is $>$\,150, namely: \ion{Fe}{I} $\lambda \lambda$ 6003.011, 6027.051, 6056.004, 6065.481, and 6252.555 {\AA}. The FFT of those lines is shown with different colors in Fig.~\ref{fft}. For each of those we computed both v $\sin i$ (first zero of the FFT) and $\Theta_{RT}$ (by goodness-of-fit technique), and the weighted means of the results are reported in Table~\ref{table_sum_spectro}. As expected the macroturbulence velocity completely dominates the rotational profile, being consistent with the value of v$_{br}$ given by \citet{trentin2022}.

Except for lithium, the abundances reported in Table~\ref{table_abund} have been calculated by \citet{trentin2022}. Regarding lithium, we proceeded with this target as described in \citet{catanzaro2020}, i.e. by a spectral synthesis performed using code \citep[SYNTHE][]{1981SAOSR.391.....K} applied to an LTE plane-parallel atmosphere model computed by ATLAS9 \citep{1993ASPC...44...87K} for T$_{eff}$ and $\log g$ reported in Table~\ref{table_sum_spectro}.
According to \citet{2020A&A...642A..62A} the non-LTE departure for [Li/H] in giant stars with the metallicity of our target, is $\approx$\,0.05\,dex, negligible for this study.
In Fig.~\ref{spectra} we show the comparison between observed and synthetic spectra in three main spectral regions, namely, H$\beta$ and H$\alpha$, and the \ion{Li}{I} 6708.766 {\AA} line. For the synthesis of the \ion{Li}{I} line, we took into account the hyperfine structure and the close \ion{Fe}{I} 6708.282 {\AA} line. The abundances of all the 24 species we detected in the spectrum of OGLE-GD-CEP-0516 are displayed in Fig.~\ref{pattern}. The analysis of this chemical pattern shows clearly that our target is metal-poor, being the most contributors to the metallicity under-abundant compared with the solar composition \citep{2010Ap&SS.328..179G}. Furthermore, since [Fe/H]\,=\,$-$0.54~dex, we can conclude that OGLE-GD-CEP-0516 is the most metal-poor lithium-rich Galactic Cepheid.

The lithium line was reproduced with an abundance A(Li)\,=\,3.06\,$\pm$\,0.10, the error on the lithium abundance was calculated by propagating the errors on the atmospheric parameters, i.e. $\delta T$, $\delta \log g$, and $\delta \xi$. This value is greater than the results from the standard Big Bang nucleosynthesis theory, which predicts a lithium abundance of A(Li)\,=\,2.72\,$\pm$\,0.06 dex \citep{2008JCAP...11..012C}, but it is in agreement with the predictions of the Galactic Chemical Evolution models \citep[GCE, e.g.][]{Romano2021}.

\section{Discussion}
\label{sect:discus}

The luminosity and the effective temperature derived in the previous section for OGLE\,GD-CEP-0516 allowed us to place the star in the HR diagram. According to the evolutionary tracks by \citet{Bressan2012} its mass is M$\approx$\,2.7\,M$_\odot$. For comparison purposes, we placed in the same HR diagram all the DCEPs with lithium discovered so far, after a new estimate of their luminosities. In particular, we used the effective temperatures from the original sources, i.e. from \citet{2019MNRAS.488.3211K} for all the stars except for V363\,Cas taken from \citet{catanzaro2020}, extinctions from \citet{2019A&A...628A..94A} and distances from \citet{Bailer2021}. All these data have been reported in Table~\ref{table_sum_phot}.

\begin{table*}
\centering
\caption{Main parameters of the known MW Li-rich DCEPs. Distances are from \citet{Bailer2021}, extinctions from \citet{2019A&A...628A..94A}, effective temperatures, lithium and iron abundances are from \citet{2019MNRAS.488.3211K} for all the stars except for V363\,Cas taken from \citet{catanzaro2020}. Luminosities have been computed by using VOSA tools.}
\label{table_sum_phot}
\begin{tabular}{lcccccr}
\hline \hline
Star & D    & A$_V$ & T$_{eff}$ & L & A(Li)& [Fe/H]~~~~~~\\
     & (pc) & (mag) & (K)        &  (L$_\odot$)     &  (dex)    &  (dex)~~~~~~     \\
\hline
ASAS\,J075842-2536.1 & 2601 $\pm$ 115  & 0.939 & 6295 & 145 $\pm$ 14  & 2.84 $\pm$ 0.10 & $-$0.16 $\pm$ 0.12\\	 
ASAS\,J131714-6605.0 & 2746 $\pm$ ~~92 & 1.686 & 6308 & 524 $\pm$ 35  & 2.96 $\pm$ 0.10 &    0.05 $\pm$ 0.06\\ 
V371\,Per            & 3255 $\pm$ 129  & 0.319 & 5973 & 474 $\pm$ 39  & 3.09 $\pm$ 0.10 & $-$0.46 $\pm$ 0.09\\  
V1033\,Cyg           & 3621 $\pm$ 374  & 2.478 & 5819 & 911 $\pm$ 190 & 3.18 $\pm$ 0.10 &    0.01 $\pm$ 0.11 \\
V363\,Cas            & 1215 $\pm$ ~~16 & 1.545 & 6660 & 287 $\pm$ 8   & 2.86 $\pm$ 0.10 & $-$0.30 $\pm$ 0.12 \\
\hline                     
\end{tabular}
\end{table*}

\begin{figure}
\centering
\includegraphics[width=\columnwidth]{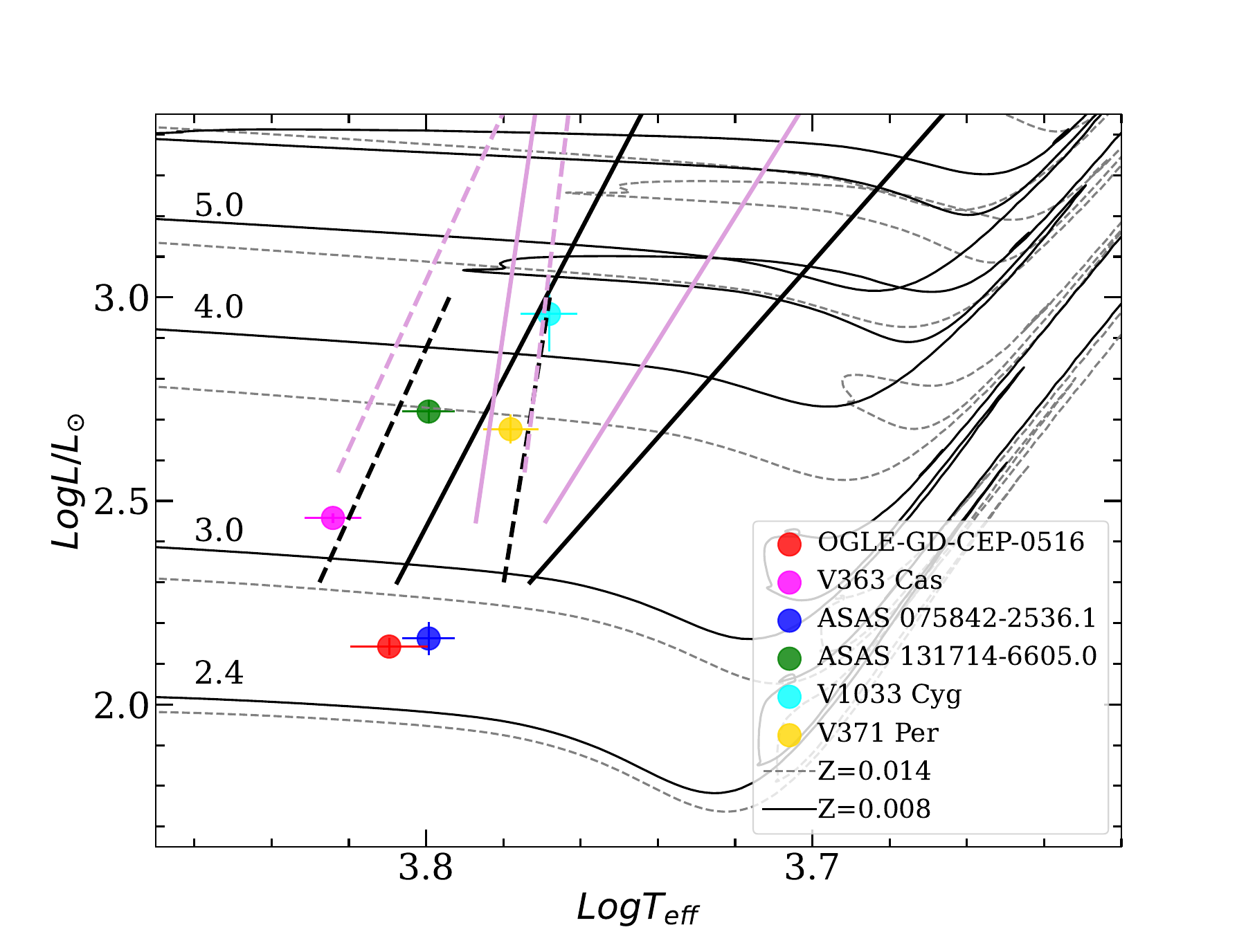}
\caption{HR diagram of the known MW Li-rich DCEPs. OGLE\,GD-CEP-0516 is shown by the filled red circle. Different symbols are used for the other five literature stars (see labels). The Instability strips for fundamental (solid lines) and first overtone (dashed lines) DCEPs by \citet{Desomma2020} as well as evolutionary tracks by \citet{Bressan2012} for Z = 0.008 (solid line) and Z = 0.014 (grey dashed line) in the mass range 2.4-–5.0 M$_{\odot}$ are also
over-plotted on the data.}
\label{HR}
\end{figure}

When the DCEPs cross the instability strip for the first time, their surface chemical composition is the same as at the end of the main sequence phase. When the first dredge-up occurs, the surface abundances of the CNO elements change because Li-free material coming from the inner part of the star is mixed into the convective envelope. In particular, carbon becomes deficient relative to its initial abundance by about $-$0.3~dex, nitrogen is increased by $\sim$0.3~dex, while oxygen should remain practically unchanged \citep[see for example,][]{2011MNRAS.410.1774L,2015MNRAS.446.3447L,2014ApJ...791...58A}. Moreover, another possible evolutionary status indicator is the abundance of Na which appears to be enhanced in post-first dredge-up intermediate-mass stars \citep{1986PASP...98..561S,1994A&A...287..113D,2013MNRAS.432..769T}. The limiting factor in this interpretation is the lack of knowledge of the initial abundance while the star is in the main sequence. In the case of OGLE\,GD-CEP-0516, we find C, O, and Na underabundant compared to the solar values, but this is not evidence for the first dredge-up because our target exhibits an overall low metallicity. This occurrence reinforces our hypothesis that the star is crossing the instability strip for the first time.

\begin{figure}
\centering
\includegraphics[width=\columnwidth]{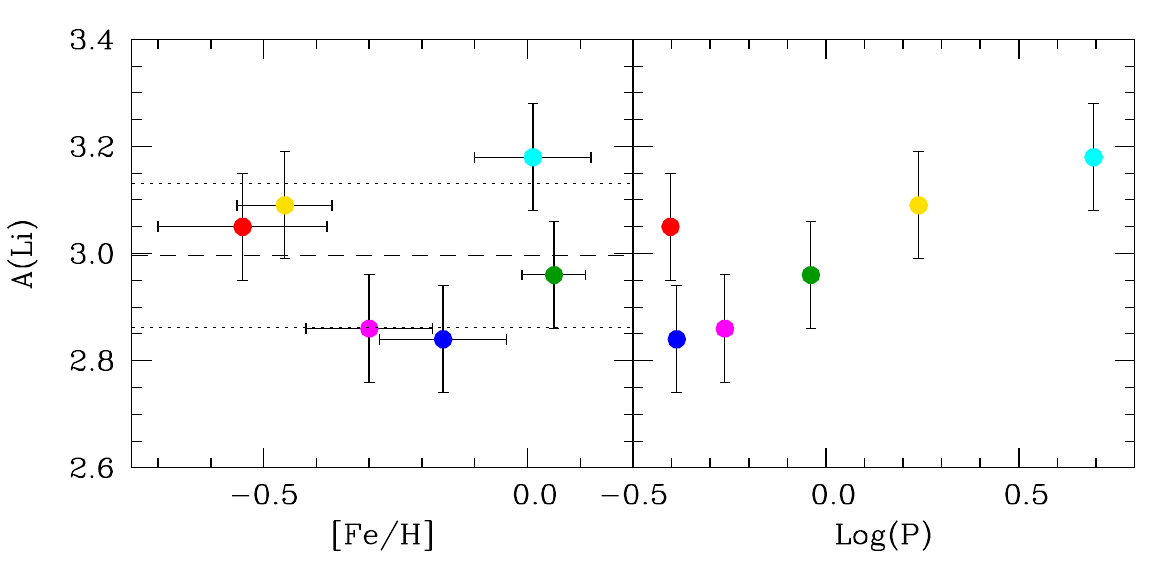}
\caption{Lithium abundances versus [Fe/H] and the logarithm of pulsation periods, respectively in the left and right panel. Colours have the same meaning as Fig.~\ref{HR}. In the left panel, the dashed line represents the average A(Li), while dotted lines represent the $\pm 1~\sigma$ level.}
\label{LiPer}
\end{figure}

\begin{figure}
\centering
\includegraphics[width=8.0cm]{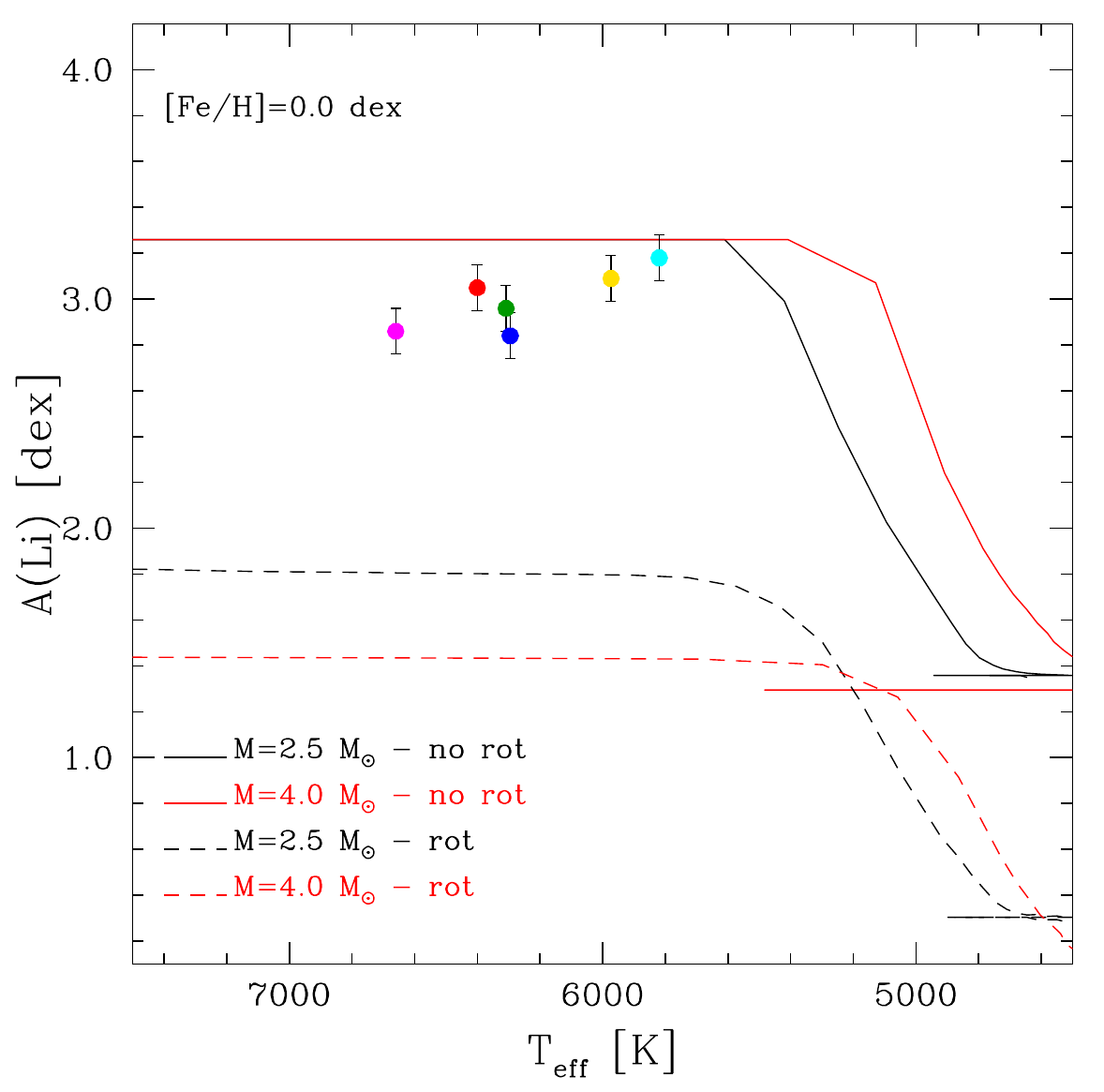}
\caption{Lithium abundances versus the effective temperatures compared with the predictions of the models STAREVOL \citep{2012A&A...543A.108L}. The curves are the predictions of the models for 2.5 M$_\odot$ and 4.0 M$_\odot$ with standard mixing (continuous lines) and with rotation-induced mixing and thermohaline instability (dashed lines).}
\label{model}
\end{figure}

To further investigate the properties of Li-rich DCEPs, we searched for possible correlations between lithium and iron abundances, and also lithium abundances versus pulsational period. Figure~\ref{LiPer} (left panel) displays an A(Li) vs [Fe/H] diagram. The sample of Li-rich DCEPs spans a relatively wide range in metallicity, from solar to [Fe/H]\,=\,-0.56~dex. The dispersion of A(Li) values around the mean is approximately 0.13~dex, exceeding the errors in individual measurements. The present sample of Li-rich DCEPs does not show any statistically significant correlation between lithium and iron. This finding is at odds with the predictions of the GCE models by e.g. \citet{Romano2021} which display a significant reduction of the lithium abundance when moving from [Fe/H]$\sim$0.0 dex to [Fe/H]$\sim -$0.5 dex (see their Fig. 6).    
The right panel of Fig~\ref{LiPer} shows the distribution of A(Li) as a function of pulsation periods. We observed a moderate to strong positive Spearman rank correlation coefficient of 0.66 between the variables. However, the associated significance of deviance from zero of 16$\%$ indicates that, although the correlation is not statistically significant at the 5$\%$ level, there is still a noteworthy relationship.  Therefore, further investigation is needed to better understand this result. As the periods of DCEPs increase with the luminosity and, in turn, thanks to the mass-luminosity relation, with the mass, we can conclude that brighter (more massive) objects show larger lithium abundances. 
To investigate more in detail this unexpected finding, in Fig.~\ref{model}, we compare the lithium abundance of the six DCEPs versus T$_{eff}$ with the predictions of the evolutionary models STAREVOL (v3.00) \citep[non-rotating and with thermohaline and rotation-induced mixings]{2012A&A...543A.108L}. More in detail, by considering models for [Fe/H]\,=\,0.0~dex, we show the evolution of A(Li) versus T$_{eff}$ for two masses, 2.5~M$_\odot$ and 4.0~M$_\odot$ both with and without rotation. According to Fig~\ref{HR} these values encompass the range of masses spanned by Li-rich DCEPs. 
Models without rotation do not predict lithium depletion within our temperature range as mass varies. Conversely, evolutionary models that take rotation into account show a strong lithium depletion of $\approx$\,1.4~dex and $\approx$\,1.8~dex, respectively for 2.5~M$_\odot$ and 4.0~M$_\odot$. According to these models, less massive objects deplete more lithium in their atmosphere if they rotate. 
These predictions are in contrast to the trend shown by Li-rich DCEPs displayed in the right panel of Fig.~\ref{LiPer}. Indeed, in the case of no rotation we should not expect any trend of lithium abundance with mass, while in the case of rotation, the expected trend is the opposite of what is observed. 
This strongly suggests that mixing due to 
rotation has not critically affected the evolution of the  
surface Li abundances in these stars, and that the measured 
A(Li) should closely reflect the initial values.
In addition, we have to take into account that the surface rotation velocity in the STAREVOL models is on the order of 80-90 km s$^{-1}$, considerably larger than what is measured for these DCEPs whose surface rotation velocity, according to our measures, should not exceed 10-20 km s$^{-1}$. This also points to a very small surface Li depletion.\\

Based on these figures, a possible working scenario able to reconcile theory and observations is to hypothesize that more massive DCEPs rotate at a slightly slower surface velocity than the less massive ones. As lithium depletion is significant for large rotation velocities, we can speculate about the difference in rotational velocity between more and less massive DCEPs that should be at most $\sim$10 km s$^{-1}$, a value compatible with the typical velocities measured in our targets.

\section{Conclusions}

In this letter, we reported the discovery of the sixth Li-rich DCEP, OGLE\,GD-CEP-0516. This object confirms the tendency of these objects to be short-period overtone pulsators. 
The presence of lithium in DCEP atmospheres is a rare event. In the course of the C-MetaLL project, we obtained spectroscopy for more than 330 DCEPs \citep[e.g.][and Trentin et al. 2024, in preparation]{trentin2022}, finding significant lithium abundance only in V363\,Cas and OGLE\,GD-CEP-0516. 
The high abundance of lithium and the low luminosity of OGLE\,GD-CEP-0516 require that it cross the IS for the first time and has not experienced the first DU. 
Considering the ensemble properties of all the six known Li-rich DCEPs in comparison with stellar evolution and GCE models, we find a general disagreement between theory and observations: i) the DCEPs span a metallicity range between 0.0 dex and $-$0.56 dex. In this interval, the measured lithium abundance is approximately constant, while GCE models predict a significant decrease to reach the plateau at A(Li)$\sim$2.7 dex; ii) the Li-rich DCEPs show larger A(Li) values at larger masses (luminosities). According to evolutionary models, this trend might be explained by hypothesizing that higher-mass DCEPs rotate slightly slower than less massive ones.

All the above conclusions are based on a small statistical sample of only six objects. An increase in the number of Li-rich DCEPs is essential to further our understanding of this puzzling phenomenon. To this aim, future wide spectroscopic surveys such as those planned with the WEAVE (WHT Enhanced Area Velocity Explorer)\footnote{https://www.ing.iac.es/astronomy/instruments/weave/weaveinst.html} and 4MOST (4-metre Multi-Object Spectroscopic Telescope)\footnote{https://www.eso.org/sci/facilities/develop/instruments/4MOST.html} may allow us to discover many new Li-rich DCEPs and obtain new insight to obtain a definitive explanation for their existence.

\begin{acknowledgements}

We thank our anonymous Referee for her/his helpful comments. This work has made use of data from the European Space Agency (ESA) mission
{\it Gaia} (\url{https://www.cosmos.esa.int/gaia}), processed by the {\it Gaia} Data Processing and Analysis Consortium (DPAC,
\url{https://www.cosmos.esa.int/web/gaia/dpac/consortium}). Funding for the DPAC has been provided by national institutions, in particular, the institutions participating in the {\it Gaia} Multilateral Agreement.
This publication makes use of VOSA, developed under the Spanish Virtual Observatory project supported by the Spanish MINECO through grant AyA2017-84089.
VOSA has been partially updated by using funding from the European Union's Horizon 2020 Research and Innovation Programme, under Grant Agreement nº 776403 (EXOPLANETS-A).
This research has made use of the SIMBAD database,
operated at CDS, Strasbourg, France.
\end{acknowledgements}

\bibliographystyle{aa}
\bibliography{cepheids}

\begin{appendix} 

\section{Table and figures}

\begin{figure*}
\centering
\includegraphics[width=13cm]{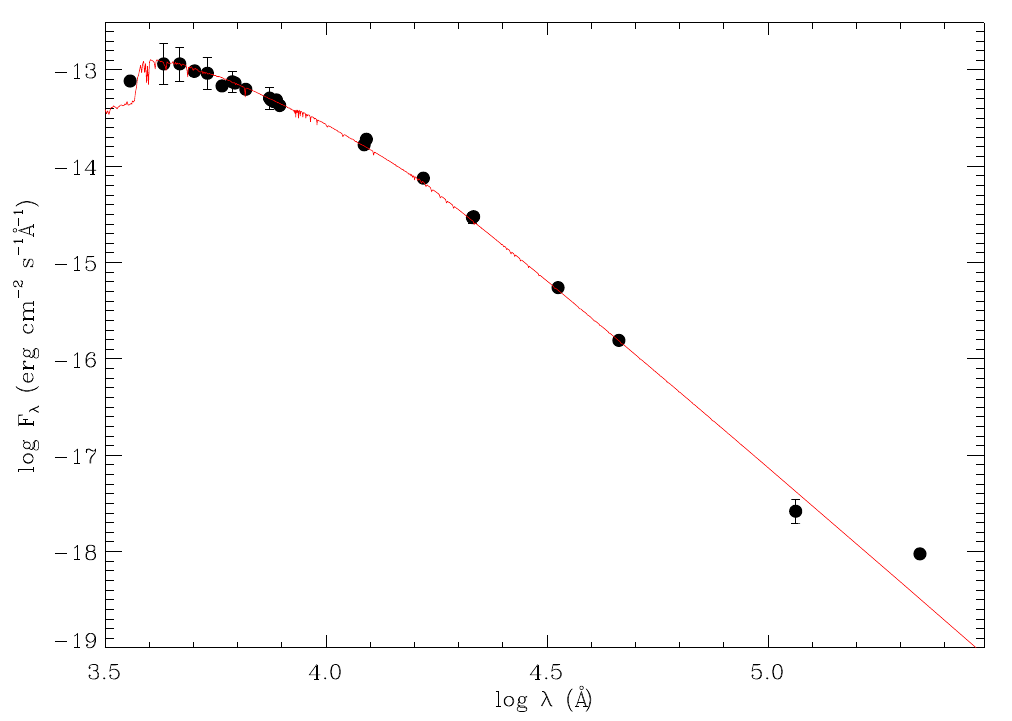}
\caption{Spectral energy distribution of OGLE-GD-CEP-0516. Filled dots represent the observed fluxes as retrieved from the VOSA tool. A red line shows the theoretical flux computed using the ATLAS9 model for T$_{\rm eff}$\,=\,6400~K and $\log g$\,=\,1.5.}
\label{sed}
\end{figure*}

\begin{figure*}
\centering
\includegraphics[width=13cm]{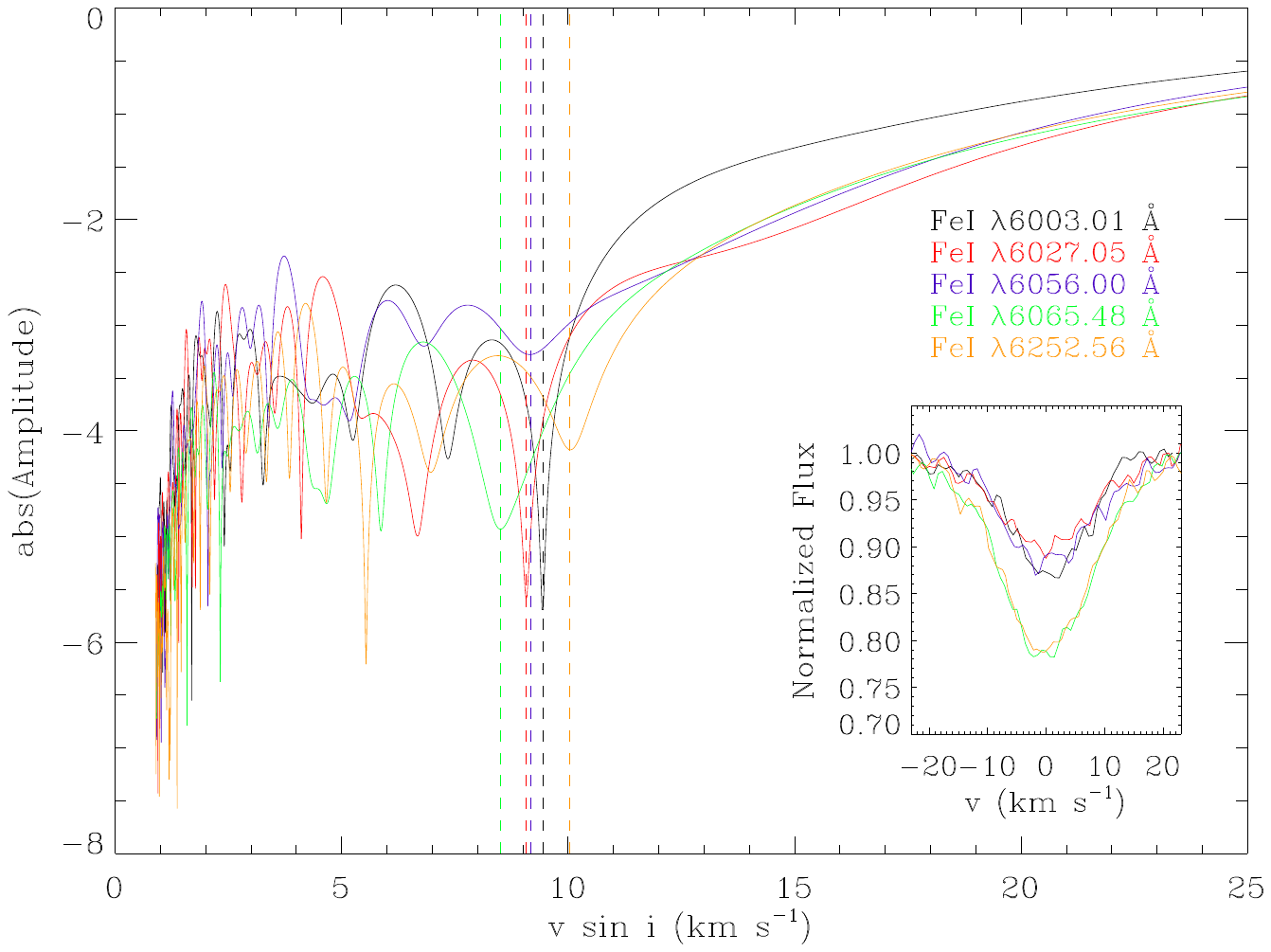}
\caption{FFT of the \ion{Fe}{i} $\lambda \lambda$ 6003.011, 6027.051, 6056.004, 6065.481, and 6252.555 {\AA}. First zeroes of the FFT, i.e. the v $\sin i$ values, are indicated by vertical dashed lines. The inset shows the spectral line profiles.}
\label{fft}
\end{figure*}

\begin{figure*}
\centering
\includegraphics[width=\textwidth]{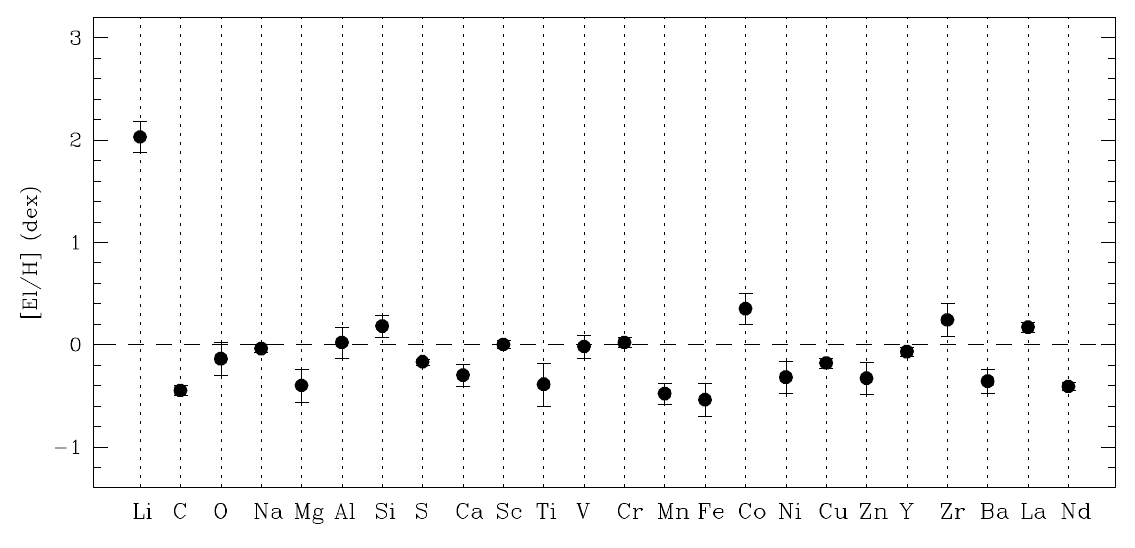}
\caption{Chemical pattern derived for our target. The dashed line represents the solar standard abundance.}
\label{pattern}
\end{figure*}

\begin{table}[h]
\centering
\caption{Elemental abundances of OGLE-GD-CEP-0516,  expressed in terms of the solar  abundances  \citep{2010Ap&SS.328..179G}, for 24 chemical species we  measured in our target. Columns labelled with N represent the number of lines used in the analysis.}
\label{table_abund}
\begin{tabular}{lrrlrr}
\hline
\hline            \noalign{\smallskip}
            
El  &  [El/H]~~~~~& N & El & [El/H]~~~~~ & N \\

            \noalign{\smallskip}
\hline
            \noalign{\smallskip}

Li   &      2.03 $\pm$ 0.10  &  1 & Cr   &       0.02 $\pm$ 0.05 & 5 \\ 
C    &   $-$0.45 $\pm$ 0.05  &  3 & Mn   &    $-$0.48 $\pm$ 0.51 & 5 \\ 
O    &   $-$0.14 $\pm$ 0.16  &  2 & Fe   &    $-$0.54 $\pm$ 0.16 &169\\ 
Na   &   $-$0.04 $\pm$ 0.04  &  6 & Co   &       0.35 $\pm$ 0.15 & 3 \\ 
Mg   &   $-$0.40 $\pm$ 0.16  &  4 & Ni   &    $-$0.32 $\pm$ 0.16 & 5 \\ 
Al   &      0.02 $\pm$ 0.15  &  2 & Cu   &    $-$0.18 $\pm$ 0.05 & 2 \\ 
Si   &      0.18 $\pm$ 0.11  &  4 & Zn   &    $-$0.33 $\pm$ 0.16 & 1 \\ 
S    &   $-$0.17 $\pm$ 0.03  &  3 & Y    &    $-$0.07 $\pm$ 0.04 & 4 \\ 
Ca   &   $-$0.30 $\pm$ 0.11  &  9 & Zr   &       0.24 $\pm$ 0.16 & 1 \\ 
Sc   &      0.00 $\pm$ 0.04  &  5 & Ba   &    $-$0.36 $\pm$ 0.12 & 4 \\ 
Ti   &   $-$0.39 $\pm$ 0.21  & 20 & La   &       0.17 $\pm$ 0.05 & 3 \\ 
V    &   $-$0.02 $\pm$ 0.11  &  1 & Nd   &    $-$0.41 $\pm$ 0.04 & 5 \\
\hline                                                                                                                                                          
\end{tabular}
\end{table}

\end{appendix}

\end{document}